\pdfoutput=1
\documentclass[aps,prl,reprint,superscriptaddress,amsmath,amssymb]{revtex4-2}
\usepackage{graphicx}
\usepackage{bm}
\usepackage[colorlinks=true,linkcolor=blue,citecolor=blue,urlcolor=blue]{hyperref}

\newcommand{\Tr}{\operatorname{tr}}
\newcommand{\Dr}[2]{D(#1\,\|\,#2)}

\begin{document}
\raggedbottom

\title{Finite-resource cooling of qudits beyond thermodynamic length}

\author{Ansh Mishra}
\email{ansh.mishra2025@gmail.com}
\affiliation{Independent researcher, Cumming, Georgia, USA}
\author{Aryan Senthilkumar}
\email{aryan.senthilk@gmail.com}
\affiliation{Independent researcher, Johns Creek, Georgia, USA}

\date{October 28, 2026}

\begin{abstract}
The heat dissipated when a quantum system is cooled with $N$ thermal machines exceeds Landauer's bound; for
swap-based collision models the excess is set by thermodynamic length. We show that for qudits the leading
finite-resource cost is instead set by a Finsler geometry generated by resonant permutations of energy eigenstates. It
improves on thermodynamic length near infinite temperature for every spectrum, and near the ground state by a factor
approaching $d-1$. To leading order coherence does not help, and at infinite temperature $d_M$-level machines gain at
most a factor $d_M-1$.
\end{abstract}

\maketitle

Landauer's principle bounds the heat dissipated when a quantum system is cooled or reset~\cite{Landauer1961,ReebWolf2014},
and the third law implies that the bound is reached only with diverging resources~\cite{Masanes2017,Wilming2017,Taranto2023}.
How the excess cost depends on the available resources has been studied in many settings, from finite-time versions of
Landauer's bound~\cite{Proesmans2020,VanVuSaito2022,Rolandi2023} to heat-bath algorithmic
cooling~\cite{Boykin2002,Baugh2005,RaeisiMosca2015,RodriguezBriones2016} and the limits of dynamic and autonomous
refrigeration~\cite{Allahverdyan2011,Silva2016,Clivaz2019}. For slowly driven processes the excess is governed by
thermodynamic length, the length of the protocol's path in a Riemannian metric on equilibrium
states~\cite{Salamon1983,Nulton1985,Crooks2007,SivakCrooks2012,Scandi2019,Abiuso2020}. Taranto \emph{et
al.}~\cite{Taranto2025} carried this picture over to Markovian collision models~\cite{Scarani2002,Ciccarello2022}, in
which the system interacts in turn with $N$ fresh thermal machines. For swaps with suitably chosen machine Hamiltonians,
the excess dissipation is $(L^*)^2/2N+O(N^{-2})$, where $L^*$ is the thermodynamic length between the initial and final
Gibbs states. They proved this optimal over all unitaries for a qubit cooled by qubit machines, and their numerics for
qutrits and ququarts suggested that it might hold in general~\cite{Taranto2025}.

Here we show that for qudits the finite-resource cost is set by a different geometry. The difference comes from
resonances: when several machine transitions are resonant with the same system transition, or system transitions share
a frequency, as in any equally spaced spectrum, a single collision can move population through several system--machine
channels at once. Since the dissipation is quadratic in the population flow, dividing the flow lowers it, much as a
current shared by two identical resistors dissipates half the heat, whereas a swap uses a single channel for each pair
of levels. We prove that, to leading order around dissipationless collisions, coherent collisions are dominated by
mixtures of resonant permutations of energy eigenstates (Theorem~1). The local cost is therefore set by a classical
Finsler geometry generated by these permutations, which reduces to thermodynamic length for qubits. It improves on
thermodynamic length near infinite temperature for every spectrum and near the ground state by a factor approaching
$d-1$, and explicit protocols save up to $50\%$ for qutrits and $83\%$ for eight-level systems. At infinite
temperature, $d_M$-level machines gain at most a factor $d_M-1$, which transfers between two levels attain
(Theorem~2). Part of the gain, near the ground state and for transfers between the extreme levels, is the advantage of
larger machines noted in Refs.~\cite{Taranto2025,Silva2016}; here it appears with machines of the same dimension as the
system, and at high temperature it comes from commensurability alone.

\emph{Setting.}---We follow Ref.~\cite{Taranto2025}. A system $S$ with Hamiltonian $H_S$ starts in the Gibbs state
$\tau_S(\beta)$ and is to be brought to $\tau_S(\lambda\beta)$, $\lambda>1$. In step $n=1,\dots,N$ it interacts through
a unitary $U_n$ with a fresh machine $M_n$, which starts in the Gibbs state $\tau_{M_n}(\beta)$ of a Hamiltonian
$H_{M_n}$ that we are free to choose. Unless stated otherwise the machines have the same dimension $d$ as the system;
Theorems~1 and~2 hold for any machine dimension $d_M$. The cost is the energy $\Delta E_M$ deposited in the machines.
Writing $\rho_{n}$ for the state of the system after step $n$, the equality form of Landauer's
principle~\cite{ReebWolf2014,Esposito2010,LandiPaternostro2021} gives
\begin{align}
\beta\Delta E_M &= \big[S(\rho_0)-S(\rho_N)\big] + \sum_{n=1}^N \Sigma_n, \label{eq:landauer}\\
\Sigma_n &= I(S\!:\!M_n)_{\rho'} + \Dr{\rho'_{M_n}}{\tau_{M_n}}, \nonumber
\end{align}
where primes denote states after the collision. The entropy difference is fixed by the endpoints, so finite resources
enter only through the dissipations $\Sigma_n\geq 0$. A swap with a machine prepared in the next state of the system,
$\tau_{M_n}=\rho_n$, dissipates $\Sigma_n=\Dr{\rho_{n-1}}{\rho_n}$, and the best swap protocol has total dissipation
$\Sigma^{\rm swap}_N=(L^*)^2/2N+O(N^{-2})$ with $L^* = 2\arccos \Tr\sqrt{\tau_S(\beta)}\sqrt{\tau_S(\lambda\beta)}$
\cite{Taranto2025}.

\emph{Dissipation in one collision.}---Each $\Sigma_n$ depends only on the incoming state $\rho=\rho_{n-1}$ of the
system, the machine state $\tau=\tau_{M_n}$ and the unitary $U=U_n$. Since $S(\rho'_{SM})=S(\rho\otimes\tau)$, it can
be written as
\begin{equation}
\Sigma = \Dr{\rho'_{SM}}{\rho\otimes\tau} - \Dr{\rho'_S}{\rho},\quad \rho'_{SM}=U(\rho\otimes\tau)U^\dagger ,
\label{eq:identity}
\end{equation}
because both sides equal $S(\rho'_S)-S(\rho\otimes\tau)-\Tr\rho'_M\log\tau$ [Supplemental Material (SM), Sec.~S1]. For
a swap with $\tau=\rho'$ the first term is $\Dr{\rho'}{\rho}+\Dr{\rho}{\rho'}$, and Eq.~\eqref{eq:identity} gives back
$\Dr{\rho}{\rho'}$.

\emph{Collision geometry.}---We first restrict to permutation unitaries; Theorem~1 below shows that, to leading order,
nothing is lost by doing so. Let $\rho=\mathrm{diag}(p)$ and $\tau=\mathrm{diag}(q)$, and let $U$ permute the product
states $|x,y\rangle$ (system level $x$, machine level $y$) according to a permutation $\pi$. The configuration weights
$r=p\otimes q$ are then permuted, and Eq.~\eqref{eq:identity} becomes $\Sigma=\Dr{r\circ\pi^{-1}}{r}-\Dr{p'}{p}$. The
system changes little if $\pi$ exchanges configurations of nearly equal weight, and we call $\pi$ \emph{resonant} at
$p$ if there is a machine state $q^*$ with $r(c)=r(\pi(c))$ for every configuration $c$, so that nothing moves.
Perturbing the machine state to $q=q^*+\delta q$ then moves the system by $\delta p=A_\pi\,\delta q$, at a cost
\begin{equation}
\Sigma=\tfrac12\Big[\sum_c \frac{(\delta r_c)^2}{r_c} - \sum_x \frac{(\delta p_x)^2}{p_x}\Big] + O(\delta q^3),
\label{eq:local}
\end{equation}
which is a quadratic form in $\delta q$ (SM, Sec.~S2). For the swap, $q^*=p$ and Eq.~\eqref{eq:local} reduces to
$\tfrac12\|\delta p\|_F^2$, where $\|\delta p\|_F^2=\sum_x(\delta p_x)^2/p_x$ is the Fisher metric; its geodesic
distance between Gibbs states is $L^*$. If a protocol uses permutations that remain resonant along a slow path $p(t)$,
its cost is, to leading order, $\tfrac{1}{2N}(\int\|\dot p\|_{\rm eff}\,dt)^2$. Here $\|\cdot\|_{\rm eff}$ is the norm
whose unit ball is the closed convex hull of the unit sets $\{\delta p:\Sigma\leq\frac12\text{ in
Eq.~\eqref{eq:local}}\}$ of all pairs $(\pi,q^*)$ resonant at $p$. In general it is a Finsler norm rather than a
Riemannian one, and it can change discontinuously across resonance manifolds. Swaps, and the permutation
$\pi_{\rm ext}$ introduced below, are resonant at every $p$, and thermodynamic length corresponds to using swaps only.

For qubits, only two resonant permutations move the system, the swap $|0,1\rangle\leftrightarrow|1,0\rangle$ and the
exchange $|0,0\rangle\leftrightarrow|1,1\rangle$, and both give the Fisher norm (SM, Sec.~S3), in agreement with
Ref.~\cite{Taranto2025}. Of the $9!$ qutrit permutations, $575$ are resonant at a generic $p$. For the equally spaced
qutrit cooled between $\beta\omega=1$ and $3$, the hull at interior points of the Fisher--Rao geodesic coincides with
the Fisher ellipse in the direction of motion, so swaps are locally optimal along that path, consistent with the
numerics of Ref.~\cite{Taranto2025}. On the resonance manifold $p_1^2=p_0p_2$, however, which contains every Gibbs
state of the equally spaced qutrit $H_S=\omega\,\mathrm{diag}(0,1,2)$, $234$ further permutations become resonant, all
of which move the system, and the hull extends beyond the Fisher ellipse.

\emph{Coherent collisions.}---We now turn to arbitrary collisions, which may create coherence between energy
eigenstates; in other settings coherence is known to contribute to entropy production and to raise the heat cost of
erasure~\cite{Santos2019,VanVuSaito2022}. A collision is dissipationless if and only if $U=(V\otimes1)U_0$, with $V$ a
unitary on $S$ and $U_0$ commuting with $\rho\otimes\tau^*$ for some machine state $\tau^*$ (SM, Sec.~S10). Since $V$
affects neither $\Sigma$ nor the spectrum, we set $V=1$. Now take a general collision close to such a one,
$\tau=\tau^*+\epsilon\Theta$ and $U=U_0e^{i\epsilon K}$, where $\Theta$ is diagonal (a rotation of the machine basis can
be absorbed into $U$) and $K$ is any Hermitian generator. Expanding Eq.~\eqref{eq:identity} gives
$\Sigma=\epsilon^2Q+O(\epsilon^3)$ with
\begin{equation}
\begin{aligned}
Q&=\tfrac12\|X\|^2_{\rho\otimes\tau^*}-\tfrac12\|\Tr_MX\|^2_{\rho},\\
X&=U_0DU_0^\dagger-D+U_0\,i[K,\rho\otimes\tau^*]\,U_0^\dagger,
\end{aligned}
\label{eq:Q}
\end{equation}
where $D=\rho\otimes\Theta$ and $\|\cdot\|_\sigma$ is the Kubo--Mori norm, the Hessian of the relative
entropy~\cite{Petz1996}.

\emph{Theorem 1.} There exist permutations $P_k$ of a product eigenbasis of $\rho\otimes\tau^*$ that commute with
$\rho\otimes\tau^*$, i.e., are resonant, and weights $\lambda_k\geq0$ with $\sum_k\lambda_k=1$, such that to first
order in $\epsilon$ the collision changes the spectrum of $\rho$ by $\sum_k\lambda_k\delta_k$, and
$Q\geq\sum_k\lambda_kQ_k$. Here $\delta_k$ and $Q_k$ are the change of the spectrum and the cost of $P_k$ with the
same machine perturbation $\Theta$.

The proof (SM, Sec.~S10) splits $X$ into a part inside the degenerate eigenspaces of $\rho\otimes\tau^*$, generated by
$\Theta$, and a part between different eigenspaces, generated by $K$. By monotonicity of the Kubo--Mori metric under
the partial trace, the coherences created by the second part cost more in the first term of Eq.~\eqref{eq:identity}
than they give back in the second. For the first part, after a rotation inside the degenerate eigenspaces of $\rho$,
only the doubly stochastic matrix $B_{ij}=|(U_0)_{ij}|^2$ matters. Birkhoff's theorem~\cite{Birkhoff1946} writes it as
$\sum_k\lambda_kP_k$, and the bound follows because the second term, $\tfrac12\|\delta p\|_F^2$, is convex in
$\delta p$. Since $Q_k\geq\tfrac12\|\delta_k\|_{\rm eff}^2$ by definition of the hull, every collision close to a
dissipationless one satisfies $\Sigma\geq\tfrac12\|\delta p\|^2_{\rm eff}+O(\epsilon^3)$.

\begin{figure}[t]
\includegraphics[width=\columnwidth]{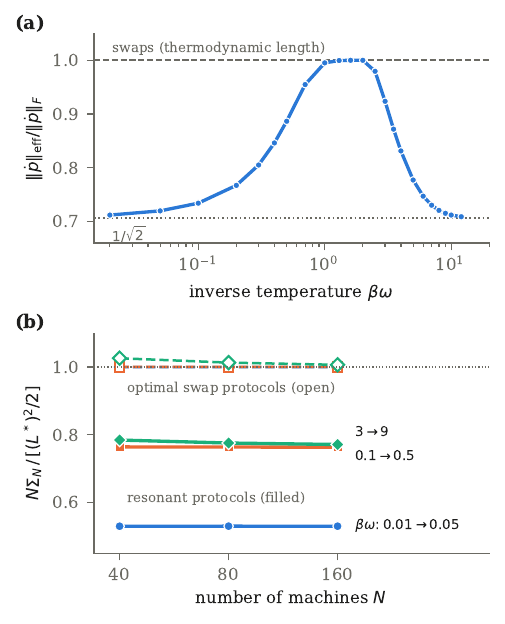}
\caption{(a) Cost per unit thermodynamic length, $\|\dot p\|_{\rm eff}/\|\dot p\|_F$, along the Gibbs curve of the
equally spaced qutrit. Here $\|\cdot\|_{\rm eff}$ is the norm whose unit ball is the convex hull of the unit sets of all
resonant permutations, computed from primal and dual cone programs that agree to within $2\times10^{-7}$. By Theorem~1
it is optimal among collisions close to dissipationless ones. Swaps give $1$. The ratio tends to $1/\sqrt2$, i.e., half
the dissipation, at high and at low temperature. (b) Exact finite-$N$ protocols: dissipation $N\Sigma_N$ in units of
the thermodynamic-length value $(L^*)^2/2$ for three cooling windows. Open symbols, optimal swap protocols (a convex
program); filled symbols, resonant permutation protocols (SM, Sec.~S5).}
\label{fig:main}
\end{figure}

For complete protocols we consider the quasistatic regime, in which every collision is an $O(1/N)$ perturbation of one
that is dissipationless for its incoming state, and all populations stay bounded away from zero. Summing over steps,
any such protocol dissipates at least $\ell^2/2N$ to leading order, where $\ell=\sum_n\|\delta p_n\|_{\rm eff}$ is the
length of its spectral path in the collision geometry. Alternating between resonant permutations attains this bound
wherever they stay resonant along the path, as swaps, $\pi_{\rm A}$ and $\pi_{\rm ext}$ below do. For qubits this
recovers, to leading order, the optimality of swaps among all unitaries~\cite{Taranto2025}. Collisions applied $O(1/N)$
away from a resonance manifold are also dominated by permutations, but the offset could enlarge the hull; in an
optimistic relaxation at six temperatures along the curve of Fig.~\ref{fig:main}(a), it does so by at most $3.6\%$
(SM, Sec.~S10).

\emph{Qutrits.}---The mechanism is most transparent at infinite temperature, where all nine configurations have
weight $1/9$ and every permutation is resonant. The Gibbs curve leaves the maximally mixed state in the direction
$\delta p\propto(1,0,-1)$, moving population from level $2$ to level $0$. A swap achieves a transfer $M$ at cost $3M^2$.
The permutation
\begin{equation}
\pi_{\rm A}:\quad |0,0\rangle\leftrightarrow|2,2\rangle,\qquad |0,1\rangle\leftrightarrow|2,0\rangle
\label{eq:piA}
\end{equation}
instead routes the transfer through two channels that use different machine transitions. A machine perturbation
$\delta q=\epsilon(0,1,-1)$ sends $m=\epsilon/3$ through each, and Eq.~\eqref{eq:local} gives
$\Sigma=9(m_1^2+m_2^2)-3(m_1+m_2)^2=\tfrac32 M^2$ for $m_1=m_2=M/2$, half the swap cost. No single permutation does
better (SM, Sec.~S4). For qutrit machines, a dual certificate over all $9!$ permutations and machine states shows that
mixtures and alternations do not help either, so by Theorem~1 half the swap cost is optimal at infinite temperature
among all qutrit-machine collisions close to dissipationless ones (SM, Sec.~S10); Theorem~2 below extends this bound to
every direction.

Away from infinite temperature, $\pi_{\rm A}$ stays resonant for every $p$, with machine populations
$q\propto(1,e^{-2\beta\omega},e^{2\beta\omega})$, i.e., an equally spaced machine with gap $2\omega$. At intermediate and
low temperatures, however, the advantage comes from permutations that are resonant only on the manifold $p_1^2=p_0p_2$
(SM, Sec.~S5). As $\beta\omega\to\infty$, a pair of $3$-cycles, $\pi_{\rm C}$, routes the transfer between the two lowest
levels through two degenerate excited levels of the machine, and an even split again halves the cost (SM, Sec.~S6).
The resulting ratio, shown in Fig.~\ref{fig:main}(a), is below $1$ except for $1.3\lesssim\beta\omega\lesssim2$.

\begin{table}[t]
\caption{Exact protocols with $41$ machines for equally spaced $d$-level systems. ``Swap'' is the optimum over swap
protocols with the same number of machines and exact endpoints, a convex problem (SM, Sec.~S5). ``Resonant'' uses the
permutations listed in SM, Secs.~S5, S7 and S9, and ends exactly in $\tau_S(\lambda\beta)$; it was evaluated in
40-digit arithmetic with explicit machine Hamiltonians $H_{M_n}=-\beta^{-1}\log q_n$. Dissipations are in units of
$10^{-6}$.}
\label{tab:exact}
\begin{ruledtabular}
\begin{tabular}{ccccc}
$d$ & $\beta\omega\to\lambda\beta\omega$ & swap & resonant & reduction \\ \hline
$3$ & $0.01\to0.05$ & $13.001$ & $6.911$ & $46.8\%$ \\
$3$ & $0.1\to0.5$ & $1235.9$ & $967.5$ & $21.7\%$ \\
$3$ & $3\to9$ & $2288.5$ & $1794.1$ & $21.6\%$ \\
$3$ & $8\to16$ & $16.288$ & $8.193$ & $49.7\%$ \\
$4$ & $0.05\to0.25$ & $596.37$ & $451.72$ & $24.3\%$ \\
$4$ & $3\to6$ & $1533.6$ & $872.2$ & $43.1\%$ \\
$4$ & $8\to16$ & $16.288$ & $6.060$ & $62.8\%$ \\
$5$ & $8\to16$ & $16.288$ & $4.613$ & $71.7\%$ \\
$6$ & $8\to16$ & $16.288$ & $3.730$ & $77.1\%$ \\
$7$ & $8\to16$ & $16.288$ & $3.137$ & $80.7\%$ \\
$8$ & $8\to16$ & $16.288$ & $2.714$ & $83.3\%$ \\
\end{tabular}
\end{ruledtabular}
\end{table}

\emph{Exact protocols.}---We also constructed complete protocols with exact dynamics. For each cooling window we fixed
a pattern of permutations and optimized all machine populations, with the final state constrained to
$\tau_S(\lambda\beta)$. A small residual ($\lesssim10^{-7}$) is removed by one extra swap, which is included in the
machine count. Each protocol was then re-evaluated in 40-digit arithmetic with explicit Hamiltonians, $d^2\times d^2$
permutation unitaries and the energy balance~\eqref{eq:landauer}. As a benchmark we minimized $\sum_n\Dr{p_{n-1}}{p_n}$
over swap protocols; this function is jointly convex, so the benchmark is the global optimum over all machine
Hamiltonians. In all cases of Table~\ref{tab:exact} the resonant protocol dissipates $22$--$83\%$ less than the best
swap protocol with the same number of machines, and the advantage persists at leading order in $1/N$
[Fig.~\ref{fig:main}(b) and SM, Sec.~S5]. The resonant machines use level spacings up to about twice those of the swap
protocols, but since the benchmark allows any machine Hamiltonian, the gain is not a matter of larger gaps.

\emph{Higher dimensions.}---For the equally spaced ququart, permutations that conserve $H_S+H_M$ for integer machine
spectra give cost ratios [as in Fig.~\ref{fig:main}(a)] between $0.72$ and $0.97$ at all five temperatures we tested
(SM, Sec.~S7), and exact protocols save $24\%$ and $43\%$ (Table~\ref{tab:exact}). For an equally spaced $d$-level
system of any dimension, the permutation
\begin{equation}
\pi_{\rm ext}:\quad |0,y\rangle\leftrightarrow|d-1,y+1\rangle,\qquad y=0,\dots,d-2,
\label{eq:piext}
\end{equation}
conserves $H_S+H_M$ for a machine with gap $(d-1)\omega$ and is resonant at every $p$. At infinite temperature it moves
population between the two extreme levels through $d-1$ parallel channels, at $1/(d-1)$ of the swap cost (SM,
Sec.~S8). The cooling direction $g_x\propto\bar E-E_x$ is not parallel to this transfer, but alternating $\pi_{\rm ext}$
with swaps realizes any convex combination of their unit sets. This beats swaps whenever the Fisher cosine between the
cooling direction and the extreme transfer exceeds $(d-1)^{-1/2}$, which holds for all $d\geq3$ (SM, Sec.~S8). Near
infinite temperature the leading coefficient is therefore at most $0.50$, $0.65$, $0.70$ and $0.73$ times the
thermodynamic-length value for $d=3$, $4$, $5$ and $6$, and with $41$ machines, exact protocols alternating swaps and
$\pi_{\rm ext}$ save $33\%$, $28\%$ and $23\%$ for $d=4$, $5$ and $6$ when cooling from $\beta\omega=0.01$ to $0.05$.

Near infinite temperature, equal spacing is unnecessary: for any spectrum $E_0<\dots<E_{d-1}$, $\pi_{\rm ext}$ is
resonant at every $p$ with a machine of level spacing $E_{d-1}-E_0$, and its cost ratio at infinite temperature is
again $1/(d-1)$. Since $\sum_x(E_x-\bar E)^2\le d(E_{d-1}-E_0)^2/4$, its Fisher cosine with the cooling direction is at
least $\sqrt{2/d}$, which exceeds $(d-1)^{-1/2}$ for every $d\geq3$ (SM, Sec.~S8). Swaps are therefore suboptimal near
infinite temperature for every spectrum, and for qutrits with unequal spacings, exact protocols cooling from
$\beta=0.01$ to $0.05$ save $16$--$31\%$.

\begin{figure}[t]
\includegraphics[width=\columnwidth]{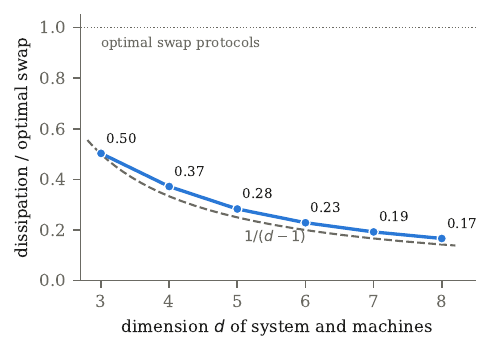}
\caption{Cooling toward the ground state ($\beta\omega=8\to16$, $41$ machines). Dissipation of the resonant protocol
divided by that of the optimal swap protocol, for equally spaced $d$-level systems and $d$-level machines (filled
symbols). Dashed line: $1/(d-1)$, the cold-limit value of the local analysis (SM, Sec.~S9). The same value is the
largest possible gain at infinite temperature (Theorem~2), and we conjecture that it bounds the gain at every
temperature.}
\label{fig:ground}
\end{figure}

\emph{Toward the ground state.}---Close to zero temperature, and for a non-degenerate first excited level, the cost is
dominated by the transfer between the two lowest levels. Consider a machine with a ground level $g$ and $d-1$
degenerate excited levels $e_j$ at the lowest gap of the system, and the $d$-cycle
$|0,e_0\rangle\to|1,g\rangle\to|0,e_1\rangle\to\dots\to|0,e_{d-2}\rangle\to|0,e_0\rangle$. Its resonance condition only
fixes $q_e/q_g=p_1/p_0$, so it is resonant at every $p$ and for every spectrum. To leading order, the cost associated
with $|1,g\rangle$ is canceled by the term $-D(\rho'_S\|\rho)$ of Eq.~\eqref{eq:identity}, while the outflow is shared
among the $d-1$ configurations $|0,e_j\rangle$. The local cost of the dominant transfer is therefore $1/(d-1)$ of the
swap value (SM, Sec.~S9; for $d=3$ the cycle is part of $\pi_{\rm C}$). Exact protocols cooling from $\beta\omega=8$ to
$16$ save $63$--$83\%$ for $d=4,\dots,8$, close to the predicted saving $1-1/(d-1)$ (Table~\ref{tab:exact} and
Fig.~\ref{fig:ground}). The construction also works for unequal spacings, saving $46\%$ for $E=(0,1,2.3)$ and $59\%$
for $E=(0,1,1.7,3.1)$.

The asymptotic optimality of the thermodynamic-length protocol suggested by the numerics of Ref.~\cite{Taranto2025}
therefore fails for every $d\geq3$ and every spectrum, even with machines of the same dimension as the system;
Theorem~1 of Ref.~\cite{Taranto2025}, which concerns swap protocols, is unaffected. At intermediate temperatures the
gain does rely on nearly commensurate levels: for $H_S=\omega\,\mathrm{diag}(0,1,2+\delta)$ and $\beta\omega=0.1\to0.5$,
the pattern $\pi_{\rm B},\pi_{\rm A}$ of SM Sec.~S5 saves $23\%$ at $\delta=0.02$ and $15\%$ at $\delta=0.1$ but nothing
at $\delta=0.3$, where $\pi_{\rm ext}$ still saves $31\%$ closer to infinite temperature.

\emph{Maximal gain.}---Our second theorem limits the possible gain.

\emph{Theorem 2.} Let the machines have $d_M$ levels. At the maximally mixed state of the system, every collision close
to a dissipationless one satisfies
\begin{equation}
\Sigma\;\geq\;\frac{1}{d_M-1}\,\tfrac12\|\delta p\|_F^2+O(\epsilon^3)
\label{eq:maxgain}
\end{equation}
for every $\delta p$. For qubit machines ($d_M=2$), Eq.~\eqref{eq:maxgain} holds at every state of the system and for
every $d$.

By Theorem~1 it suffices to consider resonant permutations. For each system level $x$, the machine levels from which
population flows into the configurations $|x,y\rangle$ define a map from the set of machine levels to itself. The flows
cancel around every cycle of this map, so at least one machine level drops out of the transfer, and the Cauchy--Schwarz
inequality gives the factor $(d_M-1)/d_M$ (SM, Sec.~S11). Transfers between two levels reach the bound: with a uniform
machine, the exchange $|0,y\rangle\leftrightarrow|d-1,y+1\rangle$, $y=0,\dots,d_M-2$, attains it for every $d$ and $d_M$
(for $d_M=d$ this is $\pi_{\rm ext}$), and the cycle used near the ground state approaches the same factor. We
conjecture that Eq.~\eqref{eq:maxgain} holds at every temperature. It holds in all our numerical tests, including
exhaustive enumeration for $(d,d_M)=(2,3)$, $(2,4)$ and $(3,3)$, about $95\,000$ permutations for $(3,4)$ and $(4,4)$,
and a search for violations over about $9\,700$ random resonant permutations (SM, Sec.~S11). If it holds, the optimal
quasistatic cost lies between $(L^*)^2/2(d_M-1)N$ and $(L^*)^2/2N$.

\emph{Discussion.}---Thermodynamic length governs swap-based cooling~\cite{Taranto2025}, but for qudits it does not set
the finite-resource limit. The leading-order cost is set instead by a Finsler geometry generated by resonant
permutations, which improves on the Fisher geometry near infinite temperature and near the ground state for every
spectrum, and for the equally spaced qutrit and ququart at most intermediate temperatures as well (for the qutrit,
everywhere outside $1.3\lesssim\beta\omega\lesssim2$). The cost still scales as $1/N$, but its prefactor is set by a
different geometry, and the gain grows with the number of levels, approaching $d-1$ near the ground state, the largest
gain that Theorem~2 allows at infinite temperature. Since the protocols only permute energy eigenstates, using machines
whose gaps match system transitions, they suit platforms with equally spaced ladders, such as Zeeman sublevels of
trapped-ion qudits~\cite{Ringbauer2022} or bosonic modes. Permutations of energy eigenstates are also the elementary
steps of heat-bath algorithmic cooling~\cite{Boykin2002,RaeisiMosca2015}, where work has mainly concerned the
achievable polarization rather than the dissipation at finite $N$; the same analysis could be applied to its
compression steps. In bosonic sub-bath cooling, exchanges of several excitations similarly improve on Gaussian
operations~\cite{Png2026}. To leading order around dissipationless collisions, Theorem~1 turns the remaining questions
into classical ones: whether Eq.~\eqref{eq:maxgain} holds at all temperatures for $d_M\geq3$, what the norm is for
$d\geq4$ and for generic spectra at low temperature, and whether near-resonant protocols can exploit their offset from
a resonance manifold.

\begin{acknowledgments}
AI tools were used in formalizing the mathematics and in drafting the paper. The authors take full responsibility for
all content and have carefully checked every proof.
\end{acknowledgments}

\clearpage
\onecolumngrid
\appendix
\section*{Supplemental Material}

\section{S1. Proof of the one-step identity}
Let $\rho'_{SM}=U(\rho\otimes\tau)U^\dagger$. Then $I(S\!:\!M)_{\rho'}+\Dr{\rho'_M}{\tau}
=S(\rho'_S)+S(\rho'_M)-S(\rho'_{SM})-S(\rho'_M)-\Tr\rho'_M\log\tau = S(\rho'_S)-S(\rho\otimes\tau)-\Tr\rho'_M\log\tau$.
On the other hand $\Dr{\rho'_{SM}}{\rho\otimes\tau}=-S(\rho\otimes\tau)-\Tr\rho'_S\log\rho-\Tr\rho'_M\log\tau$ and
$\Dr{\rho'_S}{\rho}=-S(\rho'_S)-\Tr\rho'_S\log\rho$. Subtracting gives Eq.~\eqref{eq:identity}. In the protocol, the
machine $M_n$ is uncorrelated with the system before step $n$, so its energy change is fixed by the bipartite state of
$S$ and $M_n$, which evolves from $\rho_{n-1}\otimes\tau_{M_n}$. Summing over $n$ gives Eq.~\eqref{eq:landauer} with
$\beta\Delta E_{M_n}=S(\rho_{n-1})-S(\rho_n)+\Sigma_n$.

\section{S2. Local expansion}
For a permutation $\pi$ of the configurations $c=(x,y)$ we have $r'=r\circ\pi^{-1}$ and, exactly,
$\Sigma=\Dr{r'}{r}-\Dr{p'}{p}$. If $\pi$ is resonant at $q^*$, i.e., $p_xq^*_y=p_{x'}q^*_{y'}$ whenever
$\pi(x,y)=(x',y')$, then $r^*\circ\pi^{-1}=r^*$. Write $q=q^*+\delta q$ with $\sum_y\delta q_y=0$. Then
$\delta r_{\pi(c)}=p_{x(c)}\delta q_{y(c)}-p_{x(\pi c)}\delta q_{y(\pi c)}$ and $\delta p_x=\sum_y\delta r_{(x,y)}$, both
linear in $\delta q$, and expanding both relative entropies to second order gives Eq.~\eqref{eq:local}. Resonance is a
linear condition on $\log q^*$: for each step $(x,y)\mapsto(x',y')$ of a nontrivial cycle,
$\log q^*_y-\log q^*_{y'}=\log p_{x'}-\log p_x$. Among the $9!-1$ non-identity qutrit permutations, these equations
can be solved for $575$ at a generic $p$, and $360$ of these move the system. On the manifold $p_1^2=p_0p_2$ they can
be solved for $809$, of which $594$ move the system. We enumerated all of them.

\section{S3. Qubits}
Of the $4!$ permutations of the configurations of two qubits, five admit a resonant machine for generic $p$. Three of
them only relabel machine levels at a fixed system level and do not move the system. The other two are the swap
$|0,1\rangle\leftrightarrow|1,0\rangle$, resonant at $q^*=p$, and the exchange $|0,0\rangle\leftrightarrow|1,1\rangle$,
resonant at $q^*=(p_1,p_0)$. For both, Eq.~\eqref{eq:local} equals $\tfrac12\|\delta p\|_F^2$, so the local norm is the
Fisher norm, in agreement with the qubit optimality theorem of Ref.~\cite{Taranto2025}.

\section{S4. Infinite temperature}
At $p=q=(1/d,\dots,1/d)$ every permutation is resonant and
$\delta r_{c'}=d^{-1}(\delta q_{y(\pi^{-1}c')}-\delta q_{y(c')})$. For a fixed $\pi$ and a target direction $g$,
minimizing Eq.~\eqref{eq:local} over $\delta q$ subject to $\delta p\propto g$ is a quadratic program with a
closed-form solution. For $d=3$ and $g=(1,0,-1)$ we evaluated all $9!$ permutations. Of these, $176\,256$ can move the
system along $g$, and the minimum is exactly $1/2$ of the swap cost, attained for instance by $\pi_{\rm A}$ of
Eq.~\eqref{eq:piA}. Section~S10 extends this to mixtures and alternations of permutations, to all machine states and,
via Theorem~1, to all collisions close to dissipationless ones. For $d\geq4$ we minimized over $\pi$ by simulated
annealing. For the cooling direction of the equally spaced $d$-level system, $g_x\propto\bar E-E_x$, the best
permutations found give ratios $0.567$ ($d=4$), $0.400$ ($d=5$) and $0.408$ ($d=6$); since the search is heuristic,
these are only upper bounds. For transfers between the extreme levels, $d-1$ parallel channels achieve the ratio
$1/(d-1)$.

\section{S5. Protocols and numerics}
We label configuration $(x,y)$ by $c=3x+y$ and list a permutation as $(\pi(0),\dots,\pi(8))$. The protocols use
$\pi_{\rm A}=(8,6,2,3,4,5,1,7,0)$ [Eq.~\eqref{eq:piA}], $\pi_{\rm B}=(5,3,2,8,4,0,6,7,1)$,
$\pi_{\rm C}=(5,0,2,8,3,1,6,7,4)$ and $\pi_{\rm D}=(8,5,2,7,0,1,6,3,4)$. Of these, $\pi_{\rm A}$ is resonant at every
$p$, while $\pi_{\rm B}$, $\pi_{\rm C}$ and $\pi_{\rm D}$ are resonant only on $p_1^2=p_0p_2$, where their resonance
conditions become consistent. In Table~I, the window $\beta\omega=0.01\to0.05$ uses $\pi_{\rm A}$ in every step followed
by one final swap, $0.1\to0.5$ alternates $\pi_{\rm B}$ and $\pi_{\rm A}$, $3\to9$ alternates $\pi_{\rm C}$ and
$\pi_{\rm D}$, and $8\to16$ uses $\pi_{\rm C}$ in every step.

The $N$-scans of Fig.~1(b) use the same patterns, except that for $0.01\to0.05$ we use $\pi_{\rm A}$ in nine of every
ten steps and a swap in the tenth, so that the middle level can also adjust. For this window we find
$N\Sigma_N=2.82\times10^{-4}$ at $N=40$, $80$ and $160$, against $(L^*)^2/2=5.33\times10^{-4}$; the pattern of
Table~I gives $2.76$, $2.81$ and $2.93\times10^{-4}$. For $0.1\to0.5$ we find $N\Sigma_N=0.0387$ at all three values of
$N$, against $0.0507$. For $3\to9$ we find $N\Sigma_N=0.0718$, $0.0710$ and $0.0706$, against $0.0939$, $0.0927$ and
$0.0921$ for swaps, which tend to $(L^*)^2/2=0.0915$.

Machine populations were optimized with L-BFGS on the exact dynamics, using a penalty continuation for the final state.
The swap optimum $\min\sum_n\Dr{p_{n-1}}{p_n}$ is a convex problem in the intermediate states, which we solved with
exact endpoints. We minimized over log-parametrized intermediate states, so that small populations are resolved,
starting from uniformly spaced points on the Fisher--Rao geodesic, and re-evaluated the resulting path in 40-digit
arithmetic. Where the populations are not small, the values agree with exponential-cone solutions to six or more digits,
and an independent implementation with a second solver reproduces the table entries. Penalty-based or default-tolerance
conic solutions can be inaccurate when populations are as small as $10^{-7}$, and we do not use them.

The hull ratios of Fig.~1(a) are computed from a primal--dual pair of cone programs. On the Gibbs curve $809$
permutations are resonant. For $764$ of them the resonant machine state is unique, and for the other $45$ it ranges over
a line. The primal problem is a second-order cone program over all mechanisms with a unique machine state together
with dense samples of the one-parameter families. It gives an attainable decomposition and hence an upper bound on the
gauge. Its optimal dual direction $u$ gives the lower bound $(u\cdot w)/H(u)$, where $H(u)$ is the exact support
function of the hull, with each family maximized over its whole parameter line. The two bounds agree to within
$2\times10^{-7}$ at every plotted point ($\beta\omega\leq12$). For $\beta\omega\geq6$ we use the variance form of
Sec.~S11, which remains well conditioned at small populations. At $\beta\omega=14$ the bracket is only
$[0.70773,0.70810]$, and we do not plot this point.

\section{S6. The low-temperature limit}
Let $x=e^{-\beta\omega}$, so that $p\propto(1,x,x^2)$ lies on the resonance manifold. Write $\pi_{\rm C}$ as two
$3$-cycles,
$C_1:\ |0,0\rangle\to|1,2\rangle\to|0,1\rangle\to|0,0\rangle$ and
$C_2:\ |1,0\rangle\to|2,2\rangle\to|1,1\rangle\to|1,0\rangle$.
It is resonant at $q^*\propto(x,x,1)$, a machine whose two excited levels are degenerate. For $\delta q=(a,b,-a-b)$ the
system moves by
$\delta p_0=-p_0a-p_1(a+b)$, $\delta p_2=p_1a+p_2(a+b)$ and $\delta p_1=-\delta p_0-\delta p_2$.
The configurations of $C_1$ change by
$\delta r_{|1,2\rangle}=p_0a+p_1(a+b)$, $\delta r_{|0,1\rangle}=-p_0b-p_1(a+b)$ and $\delta r_{|0,0\rangle}=p_0(b-a)$.
All three have weight $\simeq x$, while those of $C_2$ have weight $\simeq x^2$. Put $a=x\alpha$ and $b=x\gamma$. To
leading order in $x$, Eq.~\eqref{eq:local} gives
$\Sigma=\tfrac{x}{2}\left[\gamma^2+(\gamma-\alpha)^2\right]$,
while the swap cost of the same move is $\tfrac x2\alpha^2$. The minimum over $\gamma$ is at $\gamma=\alpha/2$, i.e., at
equal flow through the two channels $|0,0\rangle\to|1,2\rangle$ and $|1,2\rangle\to|0,1\rangle$, and equals
$\tfrac x4\alpha^2$, half the swap cost. The level-$2$ component of the move differs from the Gibbs tangent only at
relative order $x$. Correcting it (for instance with $\pi_{\rm D}$) costs a fraction of the total that vanishes as
$x\to0$, so $\|\dot p\|_{\rm eff}/\|\dot p\|_F\to1/\sqrt2$.

\section{S7. The convex-hull norm and ququarts}
If mechanisms $k$ are alternated over many steps, the asymptotic cost of a move $w$ is $\tfrac{1}{2N}g(w)^2$, where $g$
is the norm whose unit ball is the convex hull of the unit sets of the mechanisms (the gauge of this hull). It is the
value of the second-order cone program
\[
g(w)=\min\sum_k\|u_k\|\quad\text{subject to}\quad\sum_k D\,M_kT_k\,u_k=D\,w ,
\]
where $\delta q_k=T_ku_k$ diagonalizes the $k$-th quadratic form, $M_k$ maps $\delta q_k$ to the system move, and
$D=\mathrm{diag}(p)^{-1/2}$. For qutrits this reproduces the hull ratios of Fig.~1(a), e.g., $0.7341$ at
$\beta\omega=0.1$, $0.8313$ at $\beta\omega=4$ and $0.7207$ at $\beta\omega=8$.

For the ququart we included every permutation that conserves $H_S+H_M$ for an integer machine spectrum $e$ with
$0=e_0\le e_1\le e_2\le e_3\le 4\omega$. When a spectrum allowed more than $300$ such permutations, we sampled $300$
within its energy shells, for $8162$ mechanisms in total. The resulting ratios are upper bounds on the attainable
ratio: $0.845$, $0.894$, $0.973$, $0.957$ and $0.720$ at $\beta\omega=0.05$, $0.5$, $1$, $2$ and $4$. With
configurations labeled $c=4x+y$, the protocols of Table~I are as follows. For $\beta\omega=0.05\to0.25$ we alternate
$\pi_{\rm E}=(0,8,1,9,4,12,5,13,2,10,3,11,6,14,7,15)$, with machine spectrum $(0,2,2,4)\omega$, and
$\pi_{\rm F}=(0,4,9,13,1,5,6,7,8,12,10,14,2,3,11,15)$, with machine spectrum $(0,1,3,4)\omega$. For
$\beta\omega=3\to6$ we repeat the sequence $\pi_{\rm G},\pi_{\rm G},\pi_{\rm H}$, with
$\pi_{\rm G}=(0,2,3,4,1,6,7,8,5,10,11,12,9,15,14,13)$ [machine $(0,1,1,1)\omega$] and
$\pi_{\rm H}=(0,4,1,8,2,3,5,12,6,10,7,14,9,13,11,15)$ [machine $(0,1,1,2)\omega$]. Both protocols were re-evaluated in
40-digit arithmetic with explicit machine Hamiltonians. They reach the target to $10^{-11}$, and the swap benchmarks
are certified optimal by the convex program.

\section{S8. Every dimension $d\geq3$}
Let $p$ be uniform, so that every configuration has weight $1/d^2$, and take the linear machine perturbation
$\delta q_y=\epsilon\,(y-\tfrac{d-1}{2})$. Under $\pi_{\rm ext}$ of Eq.~\eqref{eq:piext}, each of the $d-1$
transpositions $|0,y\rangle\leftrightarrow|d-1,y+1\rangle$ moves the mass $m=(\delta q_y-\delta q_{y+1})/d=-\epsilon/d$.
The flow is therefore the same in all channels, and the system moves by $\delta p=M(e_0-e_{d-1})$ with
$M=(d-1)\epsilon/d$. Only the $2(d-1)$ exchanged configurations change, so Eq.~\eqref{eq:local} gives
\[
\Sigma=\tfrac12\Big[d^2\cdot2(d-1)\frac{\epsilon^2}{d^2}-d\cdot2M^2\Big]=\frac{(d-1)\epsilon^2}{d},
\]
whereas the swap cost of the same move is $\tfrac12\sum_x\delta p_x^2/p_x=dM^2=(d-1)^2\epsilon^2/d$. The ratio is
$1/(d-1)$. The resonance conditions $p_0q_y=p_{d-1}q_{y+1}$ can be solved for every $p$ with a geometric machine, so
$\pi_{\rm ext}$ can be used at every step without leaving its resonance manifold.

Now combine $\pi_{\rm ext}$ with swaps. In Fisher-normalized coordinates, let the cooling direction $w$ and the
extreme transfer $u=e_0-e_{d-1}$ be unit vectors at angle $\theta_d$. The attainable cost per unit length is
\[
g(w)=\min_{\epsilon\geq0}\big[a\,\epsilon+|w-\epsilon u|\big],\qquad a=(d-1)^{-1/2}=\cos\varphi_d .
\]
If $a<\cos\theta_d$ the minimum is $\cos(\theta_d-\varphi_d)<1$. At uniform $p$ the Fisher metric is $d$ times the
Euclidean one and $w\propto(\bar E-E_x)_x$, which gives
$\cos\theta_d=(d-1)/\sqrt{2\sum_x(x-\bar E)^2}=\sqrt{6(d-1)/(d(d+1))}$; the condition $a<\cos\theta_d$ is equivalent to
$5d^2-13d+6>0$. Both mechanisms are resonant at every $p$, and the local quadratic forms depend continuously on $p$.
For cooling windows close enough to infinite temperature, the leading coefficient of protocols that alternate swaps and
$\pi_{\rm ext}$ is therefore below the thermodynamic-length value $(L^*)^2/2$. For $d=3$ the construction gives
$1/\sqrt2$, the exhaustive optimum of Sec.~S4.

\emph{Arbitrary spectra.} Let $E_0<\dots<E_{d-1}$ be arbitrary, with system levels labeled by energy. The resonance
conditions of $\pi_{\rm ext}$ involve only $p_0/p_{d-1}$. They are solved by a machine whose levels are equally spaced by
$E_{d-1}-E_0$, so $\pi_{\rm ext}$ conserves $H_S+H_M$ and is resonant at every $p$. At uniform $p$ the computation above
does not involve the spectrum, so the cost ratio of the extreme transfer is again $1/(d-1)$. The cooling direction at
uniform $p$ is $w\propto(\bar E-E_x)_x$, and $\cos\theta=(E_{d-1}-E_0)/\sqrt{2\sum_x(E_x-\bar E)^2}$. Since $d$ numbers
in an interval of length $R$ satisfy $\sum_x(E_x-\bar E)^2\le dR^2/4$, we get $\cos\theta\geq\sqrt{2/d}$, which exceeds
$a=(d-1)^{-1/2}$ for all $d\geq3$ because $2(d-1)>d$. Hence swaps are beaten near infinite temperature for every
spectrum. For the qutrit, $\cos\theta=1$ for equally spaced levels, giving cost ratio $1/2$, and $\cos\theta\ge\sqrt3/2$
in general, giving a cost ratio of at most $\cos^2 15^\circ\approx0.93$. The exact protocols quoted in the main text use
$\pi_{\rm ext}$ with machine populations $q_y\propto(p_0/p_{d-1})^y$ at resonance.

\section{S9. Near the ground state, every $d$}
Let $x=p_1/p_0\to0$, with higher populations of higher order, and consider the $d$-cycle
\[
C:\ |0,e_0\rangle\to|1,g\rangle\to|0,e_1\rangle\to\dots\to|0,e_{d-2}\rangle\to|0,e_0\rangle .
\]
Take a machine with $q_{e_j}=q_e$ and $q_e/q_g=p_1/p_0$. Since $q_g\simeq1$, all $d$ configurations of $C$ then have the
same weight $w\simeq x$.

Along a cycle, a perturbation of the weights changes configuration $c_{i+1}$ by $a_{i+1}=\delta r(c_i)-\delta r(c_{i+1})$,
with $\sum_ia_i=0$. The configurations of $C$ use $d$ different machine levels, so $\delta q$ can realize any such vector.
The system moves by $\delta p_1=a_{|1,g\rangle}=:M$ and $\delta p_0=-M$.

In Eq.~\eqref{eq:local}, the term $a_{|1,g\rangle}^2/2w=M^2/2x$ is canceled at leading order by $\delta p_1^2/2p_1=M^2/2x$,
and $\delta p_0^2/2p_0=O(M^2)$ is negligible. What remains is $\sum_ja_{|0,e_j\rangle}^2/2w$ with
$\sum_ja_{|0,e_j\rangle}=-M$. Equal shares $-M/(d-1)$ minimize it, giving $\Sigma=M^2/[2x(d-1)]$. The swap cost of the
same move is $M^2/2x$ to leading order, so the ratio is $1/(d-1)$. Numerically, at $\beta\omega=14$ the ratio is
$0.5000$, $0.3333$, $0.2500$ and $0.2000$ for $d=3,\dots,6$.

The components of the cooling direction on levels $\geq2$ are smaller in Fisher norm by a factor of order
$e^{-\beta(E_2-E_1)/2}$. Correcting them with occasional swaps costs a fraction of the total that vanishes as
$\beta\to\infty$. For $d=3$ the cycle is the first $3$-cycle of $\pi_{\rm C}$.

The protocols of Table~I and Fig.~2 for $\beta\omega=8\to16$ are built as follows. For $d\geq4$ they use the $d$-cycle
in seven of every eight steps and a swap in the eighth, which adjusts the higher levels. The machine of each cycle step
starts at the resonant state $q_e/q_g=p_1/p_0$, after which all machine populations are optimized. For $d=3$ the table
uses $\pi_{\rm C}$ in every step; the cycle protocol gives a ratio of $0.545$ there. Against the swap optimum, which is
$1.6288\times10^{-5}$ for every $d$ in this window, the ratios are $0.503$, $0.372$, $0.283$, $0.229$, $0.193$ and
$0.167$ for $d=3,\dots,8$.

\section{S10. Coherent collisions: proof of Theorem 1}
Throughout, the system and machine have arbitrary finite dimensions $d$ and $d_M$, and all states are full rank. For a
state $\sigma$, the Kubo--Mori norm is $\|A\|_\sigma^2=\int_0^\infty\Tr[A(\sigma+t)^{-1}A(\sigma+t)^{-1}]\,dt$. For
diagonal $\sigma=\mathrm{diag}(s)$ it equals $\sum_{ij}|A_{ij}|^2c(s_i,s_j)$, with $c(a,b)=(\log a-\log b)/(a-b)$ and
$c(a,a)=1/a$. It is the Hessian of the relative entropy, $\Dr{\sigma+A}{\sigma+B}=\tfrac12\|A-B\|^2_\sigma+O(\|A\|^3+\|B\|^3)$
for traceless Hermitian $A,B$, and it is monotone under completely positive trace-preserving maps~\cite{Petz1996}.

\emph{Dissipationless collisions.} By Eq.~\eqref{eq:identity} and $S(\rho'_{SM})=S(\rho\otimes\tau)$, we have
$\Sigma=\Dr{\rho'_{SM}}{\rho'_S\otimes\tau}$, so $\Sigma=0$ if and only if $\rho'_{SM}=\rho'_S\otimes\tau$. In that case
$\rho\otimes\tau$ and $\rho'_S\otimes\tau$ have the same spectrum. Comparing $\Tr(\cdot)^k$ for all $k$ gives
$\Tr\rho_S'^k=\Tr\rho^k$, so $\rho'_S=V\rho V^\dagger$ for a unitary $V$, and $U_0:=(V^\dagger\otimes1)U$ commutes
with $\rho\otimes\tau$. Conversely, every such $U$ is dissipationless. A system unitary $V$ changes neither $\Sigma$
nor the spectrum of the system, so we set $V=1$.

\emph{Second order.} Let $\rho^*=\mathrm{diag}(p^*)$ and $\tau^*=\mathrm{diag}(q^*)$, put $s^*=\rho^*\otimes\tau^*$,
and let $U_0$ commute with $s^*$. To include near-resonant collisions, we allow the incoming state to differ from
$\rho^*$, $\rho=\rho^*+\epsilon R$ with $R$ diagonal, where either $R=0$ (the case of the main text, any $p^*$) or $p^*$
is non-degenerate. Let $\tau=\tau^*+\epsilon\Theta$ with $\Theta$ diagonal and traceless, and $U=U_0e^{i\epsilon K}$.
Then $\rho\otimes\tau=s^*+\epsilon D+O(\epsilon^2)$ with $D=R\otimes\tau^*+\rho^*\otimes\Theta$, and
\[
U(\rho\otimes\tau)U^\dagger-\rho\otimes\tau=\epsilon X+O(\epsilon^2),\qquad
X=\underbrace{U_0DU_0^\dagger-D}_{X_\parallel}+\underbrace{U_0\,i[K,s^*]\,U_0^\dagger}_{X_\perp}.
\]
Expanding both relative entropies of Eq.~\eqref{eq:identity} around $s^*$ and $\rho^*$ gives
$\Sigma=\epsilon^2Q+O(\epsilon^3)$ with $Q=\tfrac12\|X\|^2_{s^*}-\tfrac12\|\Tr_MX\|^2_{\rho^*}$, which is
Eq.~\eqref{eq:Q} for $R=0$.

\emph{Orthogonal splitting.} Group the product basis $|x,y\rangle$ into the eigenspaces (blocks) of $s^*$. The matrix
$i[K,s^*]$ has entries $iK_{ij}(s^*_j-s^*_i)$, which vanish inside blocks. Since $U_0$ is block diagonal, $X_\parallel$
lives inside the blocks and $X_\perp$ between them. For diagonal $s^*$ the Kubo--Mori norm is a weighted sum over
entries, so $\|X\|^2_{s^*}=\|X_\parallel\|^2_{s^*}+\|X_\perp\|^2_{s^*}$. An entry of $X_\parallel$ joining $|x,y\rangle$
and $|x',y\rangle$ requires $p^*_xq^*_y=p^*_{x'}q^*_y$, i.e., $p^*_x=p^*_{x'}$. Hence $\Tr_MX_\parallel$ lives inside
the eigenspaces of $\rho^*$. Conversely, $\Tr_MX_\perp$ has no entries there, because $|x,y\rangle$ and $|x',y\rangle$
with $p^*_x=p^*_{x'}$ lie in one block. So $Q=Q_\parallel+Q_\perp$, where
$Q_\bullet=\tfrac12\|X_\bullet\|^2_{s^*}-\tfrac12\|\Tr_MX_\bullet\|^2_{\rho^*}$.

\emph{(i) Coherent part.} $Q_\perp\geq0$, by monotonicity of the Kubo--Mori metric under the partial trace, which maps
$s^*$ to $\rho^*$.

\emph{(ii) Block part.} Conjugating by $W\otimes1$, with $W$ unitary inside the eigenspaces of $\rho^*$ (this commutes
with $s^*$ and with $D$, since $R=0$ whenever $p^*$ is degenerate), we may assume that $\Tr_MX_\parallel$ is diagonal.
Inside a block with eigenvalue $r$, $\|\cdot\|^2_{s^*}=r^{-1}\|\cdot\|^2_{\rm HS}$. For diagonal $D$,
$\|U_0DU_0^\dagger-D\|^2_{\rm HS}=2\Tr D^2-2\sum_{ij}B_{ij}D_iD_j$, where $B_{ij}=|(U_0)_{ij}|^2$ is doubly stochastic and
block diagonal. By Birkhoff's theorem~\cite{Birkhoff1946}, $B=\sum_k\lambda_kP_k$, a convex combination of block-diagonal
permutation matrices. Each $P_k$ commutes with $s^*$, and
$\|X_\parallel\|^2_{s^*}=\sum_k\lambda_k\|P_kDP_k^{\rm T}-D\|^2_{s^*}$. The diagonal of $U_0DU_0^\dagger$ is
$B\,\mathrm{diag}(D)$, so $\Tr_MX_\parallel=\mathrm{diag}(\bar\delta)$ with $\bar\delta=\sum_k\lambda_k\delta_k$ and
$\delta_k=\Tr_M(P_kDP_k^{\rm T}-D)$. By convexity of $\delta\mapsto\sum_x\delta_x^2/p^*_x$,
\[
Q_\parallel=\tfrac12\sum_k\lambda_k\|P_kDP_k^{\rm T}-D\|^2_{s^*}-\tfrac12\|\bar\delta\|^2_{\rho^*}
\;\geq\;\sum_k\lambda_kQ_k,\qquad Q_k=\tfrac12\|P_kDP_k^{\rm T}-D\|^2_{s^*}-\tfrac12\|\delta_k\|^2_{\rho^*}.
\]
For $R=0$, $Q_k$ is the cost~\eqref{eq:local} of $P_k$ with machine perturbation $\Theta$. Equality holds if and only if
all $\delta_k$ coincide.

\emph{(iii) Spectral move.} To first order, the eigenvalues of $\rho'_S=\rho+\epsilon\Tr_MX+O(\epsilon^2)$ are those of
$\rho^*+\epsilon(R+\Tr_MX)$ compressed to the eigenspaces of $\rho^*$. Since $\Tr_MX_\perp$ does not contribute, the
spectrum moves by $\epsilon\bar\delta$. Together, (i)--(iii) prove Theorem~1, including for near-resonant collisions.

\emph{Consequences.} For $R=0$, each $P_k$ with perturbation $\Theta$ is one of the mechanisms whose unit sets define
$\|\cdot\|_{\rm eff}$, so $Q_k\geq\tfrac12\|\delta_k\|^2_{\rm eff}$ and
\[
Q\;\geq\;\sum_k\lambda_kQ_k\;\geq\;\tfrac12\sum_k\lambda_k\|\delta_k\|_{\rm eff}^2\;\geq\;\tfrac12\Big(\sum_k\lambda_k\|\delta_k\|_{\rm eff}\Big)^2\;\geq\;\tfrac12\|\bar\delta\|^2_{\rm eff}.
\]
For the quasistatic regime, assume that every collision is an $O(1/N)$ perturbation of a collision that is
dissipationless for its incoming state, that system and machine populations are bounded away from zero, and that $K$
and $\Theta$ are uniformly bounded, so that the remainders are uniform. Summing over steps and using Cauchy--Schwarz
gives $\sum_n\Sigma_n\geq\tfrac12\sum_n\|\delta p_n\|^2_{\rm eff}-O(N^{-2})\geq\ell^2/2N-O(N^{-2})$, with
$\ell=\sum_n\|\delta p_n\|_{\rm eff}$. Conversely, suppose that permutations $P_k$ remain resonant along a path, with
resonant machines depending continuously on $p$. Using each $P_k$ in a fraction $\lambda_k$ of the steps then reproduces
the first-order moves and the cost $\sum_k\lambda_kQ_k$, without the Jensen gap of (ii). Taking the $P_k$, weights and
moves from an optimal decomposition of the hull gauge attains $\tfrac12\|\delta p\|^2_{\rm eff}$ per step on average.
The permutations $P_k$ act on the product basis after the rotation $W$ of (ii), i.e., on an eigenbasis of $\rho^*$
rotated within its degenerate eigenspaces.

\emph{Numerical checks.} We drew random dissipationless collisions for $(d,d_M)\in\{(3,3),(4,4),(3,4),(4,3),(3,2)\}$,
with resonances of swap, extreme-level ($\pi_{\rm ext}$-type), partial and uniform-machine type, including degenerate
system spectra, and with random block unitaries $U_0$ and random $K$ and $\Theta$, at least $150$ draws per dimension
pair. In double precision, the identity
$Q_\parallel=\sum_k\lambda_kQ_k+\tfrac12\big(\sum_k\lambda_k\|\delta_k\|^2_{\rho^*}-\|\bar\delta\|^2_{\rho^*}\big)$ held to
relative accuracy $3\times10^{-14}$, $Q_\perp$ was nonnegative in every draw, and the Birkhoff move
$\sum_k\lambda_k\delta_k$ agreed with the first-order move computed from $X$ to $3\times10^{-14}$. We also compared
Eq.~\eqref{eq:Q} with the exact dissipation in 60-digit arithmetic, using exactly resonant machines and Richardson
extrapolation in $\epsilon$, for $18$ draws with $(d,d_M)=(4,4),(3,4),(4,3),(3,3)$ that cover swap, extreme-level and
partial resonances and degenerate system spectra. The two agreed to at least $14$ significant digits, with machine
populations as small as $10^{-16}$. For the near-resonant case $R\neq0$, nine draws in 50-digit arithmetic agreed with
the formula to $10^{-21}$, satisfied $Q\geq\sum_k\lambda_kQ_k$, and had first-order moves equal to the Birkhoff mixture.

\emph{Near-resonant offsets.} The qutrit Gibbs curve lies on the resonance manifold $p_1^2=p_0p_2$, and a protocol may
stay $O(1/N)$ away from it. A permutation that is resonant only on the manifold then acts with an offset $R$ along its
normal. The offset enters $D=R\otimes\tau^*+\rho^*\otimes\Theta$ and can unlock moves that are blocked on the manifold.
To bound the possible gain, we let every mechanism choose its own offset, which is an optimistic relaxation. The hull
ratio of Fig.~1(a) then decreases as follows:
\begin{center}
\begin{tabular}{lcccccc}
$\beta\omega$ & $0.02$ & $0.1$ & $0.5$ & $1.6$ & $4$ & $8$\\ \hline
exactly resonant & $0.7120$ & $0.7341$ & $0.8866$ & $1$ & $0.8313$ & $0.7207$\\
relaxed offsets & $0.7110$ & $0.7297$ & $0.8778$ & $1$ & $0.8014$ & $0.7099$
\end{tabular}
\end{center}
Whether actual protocols can realize any of this is open. In all cases the relaxed values stay above the conjectured
bound $1/\sqrt2$.

\emph{Infinite temperature: a dual certificate.} Let $p=(\tfrac13,\tfrac13,\tfrac13)$ and $g=(1,0,-1)$, the cooling
direction of the equally spaced qutrit. Let $h$ be the largest value of $\langle g,\delta\rangle_F/\sqrt{2Q}$ over all
mechanisms and perturbations, where $\langle g,\delta\rangle_F=\sum_xg_x\delta_x/p_x$. The linear functional
$\langle g,\cdot\rangle_F/h$ is then at most $1$ on the hull, so $\|g\|_{\rm eff}\geq\|g\|_F^2/h$. For a single
mechanism, $h^2=b^{\rm T}A^+b$, where $2Q=\Theta^{\rm T}A\,\Theta$ and $\langle g,\delta\rangle_F=b^{\rm T}\Theta$; it is
infinite if $A$ has a null vector with $b^{\rm T}\Theta\neq0$. Three cases occur. (i) For a uniform machine, $q^*=p$,
all $9!$ permutations are resonant. We evaluated $h^2$ for each of them, in floating point and, independently, in exact
rational arithmetic. None is infinite. The largest value is exactly $2\|g\|_F^2=12$, attained by $1296$ permutations
including $\pi_{\rm A}$, and the next value is $15/2=1.25\|g\|_F^2$. (ii) If exactly two machine populations are equal,
say $q^*_0=q^*_1=a<\tfrac12$ after relabeling, only exchanges between machine levels $0$ and $1$ move the system. Each
changes a configuration by $\pm t$, with $t=(\theta_1-\theta_0)/3$. The move is $\delta=t\,v$ with $v_x\in\{-1,0,1\}$,
and $2Q=(6k/a-3|v|^2)t^2$, where $k\geq1$ is the number of exchanges in each direction. Since $(v_0-v_2)^2\leq4$ and
$|v|^2\leq2\leq2k$, we get $h^2\leq36/(6/a-6)=6a/(1-a)<6=\|g\|_F^2$. (iii) If the three machine populations are
distinct, no resonant permutation moves the system. Therefore $\|g\|_{\rm eff}\geq\|g\|_F/\sqrt2$, and $\pi_{\rm A}$
attains this value. With qutrit machines at the maximally mixed state, half the swap cost is thus the optimum over the
hull, and by Theorem~1 also over all collisions close to dissipationless ones. With $d_M$-level machines the optimum is
at least $1/(d_M-1)$ of the swap cost (Theorem~2); for $d_M\geq4$, three or more $\pi_{\rm ext}$-type channels can go
below $\tfrac12$.

\section{S11. Maximal gain: proof of Theorem 2}
Let $P$ be a permutation of the configurations that commutes with $s^*=\rho^*\otimes\tau^*$, and let $\Theta$ be
diagonal and traceless. Write $\kappa(c)$ for the configuration whose population flows into $c$, so that
$\kappa=P^{-1}$ and $r^*_{\kappa(c)}=r^*_c$. With $D=\rho^*\otimes\Theta$ we have $X_c=D_{\kappa(c)}-D_c$. For
$c=(x,y)$ and $\kappa(c)=(x',y')$, resonance gives $p^*_{x'}q^*_{y'}=p^*_xq^*_y$, so
\[
X_{(x,y)}=r^*_{(x,y)}\big[\psi(y')-\psi(y)\big],\qquad \psi=\Theta/q^*,\qquad \textstyle\sum_yq^*_y\psi(y)=\Tr\Theta=0 .
\]
For each system level $x$ let $\sigma_x(y)$ be the machine level of $\kappa(x,y)$, and put
$\Delta_x(y)=\psi(\sigma_x(y))-\psi(y)$. Then $\delta_x=p^*_x\,\mathbb E_{q^*}\Delta_x$. With expectations over $y\sim q^*$,
\[
2Q=\sum_xp^*_x\Big[\mathbb E(\Delta_x^2)-(\mathbb E\Delta_x)^2\Big],\qquad
\|\delta\|_F^2=\sum_xp^*_x(\mathbb E\Delta_x)^2 .
\]
Hence $\|\delta\|_F^2\leq(d_M-1)\,2Q$ is equivalent to
\begin{equation}
d_M\sum_xp^*_x(\mathbb E\Delta_x)^2\;\leq\;(d_M-1)\sum_xp^*_x\,\mathbb E(\Delta_x^2).
\tag{S11.1}
\end{equation}

\emph{(a) Maximally mixed system.} Here $p^*_x=1/d$, and resonance forces $q^*_{\sigma_x(y)}=q^*_y$. So $\sigma_x$
maps each class $\mathcal K$ of machine levels with a common population $q_{\mathcal K}$ into itself. Choose a cycle
$C_{\mathcal K}\subseteq\mathcal K$ of $\sigma_x$ in each class. Along a cycle the differences $\Delta_x$ telescope to
zero, and $q^*$ is constant on it, so the cycles do not contribute to $\mathbb E\Delta_x$. By Cauchy--Schwarz over the
remaining levels $S_x$,
\[
(\mathbb E\Delta_x)^2\leq\Big(\sum_{y\in S_x}q^*_y\Big)\,\mathbb E(\Delta_x^2)
\leq\Big(1-\sum_{\mathcal K}q_{\mathcal K}\Big)\,\mathbb E(\Delta_x^2)\leq\Big(1-\frac1{d_M}\Big)\,\mathbb E(\Delta_x^2),
\]
since $1=\sum_{\mathcal K}|\mathcal K|\,q_{\mathcal K}\leq d_M\sum_{\mathcal K}q_{\mathcal K}$. Summing over $x$
gives~(S11.1).

\emph{(b) Qubit machines, any system state.} For $d_M=2$, write $\psi=a\,(q^*_1,-q^*_0)$. Each $\sigma_x$ is one of the
four maps of $\{0,1\}$ into itself:
\begin{center}
\begin{tabular}{lcc}
map & $(\mathbb E\Delta_x)^2/a^2$ & $\mathbb E(\Delta_x^2)/a^2$\\ \hline
identity & $0$ & $0$\\
exchange & $(q^*_1-q^*_0)^2$ & $1$\\
constant $0$ & $q_1^{*2}$ & $q^*_1$\\
constant $1$ & $q_0^{*2}$ & $q^*_0$
\end{tabular}
\end{center}
Let $P_{\rm id}$, $P_{\rm ex}$, $P_0$ and $P_1$ be the total $p^*$-weights of the four types. The inflowing
configurations $\kappa(c)$ are again distributed as $r^*$, so the machine level $\sigma_X(Y)$ of $(X,Y)\sim r^*$ has
law $q^*$:
\[
P_{\rm id}q^*_1+P_{\rm ex}q^*_0+P_1=q^*_1,\qquad\text{equivalently}\qquad
P_{\rm ex}(q^*_0-q^*_1)+P_1q^*_0=P_0q^*_1 .
\]
Substituting this gives
\[
\sum_xp^*_x(\mathbb E\Delta_x)^2=a^2\big[P_{\rm ex}q^*_0(q^*_0-q^*_1)+P_1q^*_0\big],\qquad
\sum_xp^*_x\mathbb E(\Delta_x^2)=2a^2(P_{\rm ex}+P_1)q^*_0 .
\]
Then (S11.1) reads $P_{\rm ex}q^*_0(q^*_0-q^*_1)\leq P_{\rm ex}q^*_0$, which always holds.

\emph{(c) From permutations to collisions.} By Theorem~1 (with $R=0$, for any $p^*$) and convexity,
$Q\geq\sum_k\lambda_kQ_k\geq\frac{1}{d_M-1}\,\frac12\sum_k\lambda_k\|\delta_k\|_F^2\geq\frac{1}{d_M-1}\,\frac12\|\bar\delta\|_F^2$.

\emph{Attainment.} For any $d$ and $d_M$, take the maximally mixed system, a uniform machine and the exchange
$|0,y\rangle\leftrightarrow|d-1,y+1\rangle$ for $y=0,\dots,d_M-2$, with $\theta_y=\epsilon\,(y-\frac{d_M-1}{2})$. Each
exchanged configuration changes by $\pm\epsilon/d$, and the system moves by $\delta_0=-\delta_{d-1}=(d_M-1)\epsilon/d$.
Then $\sum_cX_c^2/r_c=2(d_M-1)d_M\epsilon^2/d$ and $\|\delta\|_F^2=2(d_M-1)^2\epsilon^2/d$, so
$\|\delta\|_F^2/2Q=d_M-1$, which is equality in Eq.~\eqref{eq:maxgain}. For $d_M=d$ this exchange is $\pi_{\rm ext}$
(Sec.~S8).

\emph{No free moves.} For an exactly resonant permutation, $2Q=0$ forces $\delta=0$, so $\max_\Theta\|\delta\|_F^2/2Q$
is finite at every state. Indeed, $2Q=0$ means that each $\Delta_x$ is a constant $c_x$, i.e.,
$\psi(\sigma_x(y))=\psi(y)+c_x$. Take $X\sim p^*$ independent of $Y\sim q^*$. The marginal condition
$\sigma_X(Y)\sim q^*$ then gives
$\mathbb E\,\psi(Y)^2=\mathbb E\,\psi(\sigma_X(Y))^2=\mathbb E\,[\psi(Y)+c_X]^2=\mathbb E\,\psi(Y)^2+\mathbb E\,c_X^2$,
using $\mathbb E\,\psi(Y)=0$. Hence all $c_x=0$ and $\delta=0$.

\emph{(d) Numerical evidence at general temperature.} We tested Eq.~(S11.1) by running over all permutations of the
$dd_M$ configurations. At each test state we solved the resonance equations, sampled the resonance families, and
computed $\max_\Theta\|\delta\|_F^2/2Q$. The test states were the maximally mixed state, Gibbs states of the equally
spaced system at several temperatures, Gibbs states of random spectra, and random states with all configuration weights
above $10^{-9}$. (At the maximally mixed state the enumeration is incomplete, but that case is covered by (a).) For
$(d,d_M)=(2,3)$ and $(2,4)$ the value $d_M-1$ is attained at the maximally mixed state, while elsewhere the maxima stay
below it and approach it toward the ground state ($1.985$ and $2.942$ at $\beta\omega=6$). For $(3,3)$ the maxima stay
below $2$, the largest being $1.99996$ at $\beta\omega=6$. The cases $(3,2)$ and $(4,2)$, covered by (b), serve as
checks and give at most $1$. One double-precision run at $\beta\omega=12$ returned an infinite value. This is an artifact
of configuration weights below $10^{-10}$, since by the lemma above infinite values cannot occur.

We repeated the $(3,3)$ test along the Gibbs curve in 60-digit arithmetic, with exactly resonant machines, over all
$594$ resonant permutations that move the system. No infinite value occurs, and the largest value lies below the bound
$2$ by
\begin{center}
\begin{tabular}{lcccccc}
$\beta\omega$ & $0.05$ & $6$ & $12$ & $16$ & $24$ & $40$\\ \hline
$2-\max\lambda$ & $1.2\times10^{-3}$ & $3.7\times10^{-5}$ & $2.3\times10^{-10}$ & $7.6\times10^{-14}$ &
$8.6\times10^{-21}$ & $1.1\times10^{-34}$
\end{tabular}
\end{center}
The bound is thus approached from below at both ends of the Gibbs curve. For $(d,d_M)=(3,4)$ and $(4,4)$ exhaustive
enumeration is out of reach. There we took all permutations that conserve $H_S+H_M$ for integer machine spectra up to
$4\omega$, about $5\,300$ and $9\,500$ per temperature, at $\beta\omega=0.05$, $0.5$, $2$ and $5$. The largest values
are $2.9989$ and $2.99999$, below the bound $3$, and the cases flagged in double precision were confirmed in 50-digit
arithmetic. Repeating this for random non-degenerate integer system spectra and temperatures, with about $35\,500$
mechanisms for $d=3,4$, gives a largest value of $2.995$.

The most demanding test is a direct search for violations. For a fixed permutation, the resonance conditions are linear in
$(\log p,\log q^*)$. For each of $(d,d_M)=(2,4),(2,5),(3,3),(3,4),(4,3),(4,4)$ we drew $3000$ random sparse
permutations, products of up to four short cycles. For each of the $9\,687$ that are resonant on a nontrivial set, we
searched (Nelder--Mead from three starting points) for the largest $\max_\Theta\|\delta\|_F^2/2Q$ over all system and
machine states at which the permutation is resonant. The maxima are $3.00000000$, $2.99999998$, $2.00000000$,
$2.99999997$, $2.00000000$ and $2.9995$, against bounds $3$, $4$, $2$, $3$, $2$ and $3$. The bound is never exceeded,
and it is reached for $(2,4)$, $(3,3)$, $(3,4)$ and $(4,3)$; for $(2,5)$ the sample missed the four-channel exchange
that attains $4$.

\emph{The marginal condition alone does not suffice.} Part (b) uses only the marginal condition $\sigma_X(Y)\sim q^*$,
so one may ask whether arbitrary maps $\sigma_x$ subject to this condition obey~(S11.1). For fixed $q^*$ and $\psi$, the
worst family of maps maximizes $\sum_xp_x(\mathbb E\Delta_x)^2/\sum_xp_x\mathbb E(\Delta_x^2)$, a linear-fractional
program over distributions on all $d_M^{d_M}$ maps. We maximized it over $q^*$ and $\psi$ and then recomputed the
optimal vertices in 50-digit arithmetic. For $d_M=3$ the ratio approaches $\frac23$ only from below; the largest value,
at the boundary of the simplex, is $\frac23-2.9\times10^{-10}$. For $d_M=4$, however, it exceeds $\frac34$. For
example, $q^*\approx(0.4386,0.0878,0.1782,0.2954)$ with four maps of weights $0.73$, $0.17$, $0.04$ and $0.06$ gives
$0.7553$, with the marginal condition satisfied to $10^{-50}$. These families violate weight preservation, so they do not
come from resonant permutations. For $d_M=2$ the LP-dual certificate $g=\frac12\psi^2$, which satisfies
$(\mathbb E\,\psi\circ\sigma)^2\leq\mathbb E\big[\frac12(\psi\circ\sigma-\psi)^2+g\circ\sigma-g\big]$ for every map
$\sigma$, gives a second proof of (b). For $d_M\geq4$, a proof of the general case must therefore use the full resonance
structure (bijectivity and weight preservation), not only the marginal condition. Random sampling misses these
counterexamples, which shows how special they are.

\section{S12. Code and data}
All numerical results in the Letter and in this Supplemental Material were obtained with Python scripts, which are
available together with their output logs and the protocol data at
\url{https://github.com/anshM123/Resonant-Collision-Cooling}. The README file there lists the script behind each
statement.

\end{document}